\documentclass{article}
\usepackage{amssymb, amsmath, epsf, epsfig, graphicx}
\title{The influence of spin-dependent phases of tunneling electrons on
the conductance of a point  ferromagnet/isolator/superconductor
contact}  
\author{B.P. Vodopyanov \vspace{3mm}\\Kazan Physical-Technical Institute RAS, 420029
Kazan, Russia \vspace{2mm} \\ e-mail:  vodopyanov@kfti.knc.ru}

\date{ March 17, 2008}

\begin{document}
\maketitle {\large \textbf{Abstract.}  The Andreev reflection
probability for a ferromagnet/isolator/superconductor (FIS)
contact at the arbitrary spin-dependent amplitudes of the electron
waves transmitted through and reflected from the potential barrier
is found. It is shown that Andreev reflection probabilities of
electron and hole excitations in the FIS contact are different.
The energy levels of Andreev bound states are found. The ballistic
conductance of the point FIS contact is calculated.\vspace{6mm}

PACS: {74.50.+r, 74.80.-g, 75.30.Et}\vspace{6mm}

One of the manifestations of the exchange field in a ferromagnetic
metal (F) is the presence of electron spin subbands with different
values of Fermi momenta: $p_{\,\uparrow}$  for the subband with
spin up and $p_{\,\downarrow}$  for the subband with spin down. As
a consequence, for layered F/S structures (S stands for
superconductor)\, the spatial dependence of the anomalous Green's
function (GF)  in a ferromagnet has an oscillatory character. One
of the impressive manifestations of such oscillations and related
phase shifts is a recent observation of spontaneous zero-field
supercurrents at temperature lower than the junction $0-\pi$
transition temperature in superconducting networks of SFS
junctions with weakly ferromagnetic  barriers \cite{Ryazanov}. The
influence of the oscillatory character of the anomalous GF in a
ferromagnet on the properties of various hybrid F/S structures is
studied well enough (see reviews [2-4]).

Another consequence is the suppression of Andreev reflection
\cite{Andreev}. When a polarized electron from the subband with,
for example, spin up gets into a superconductor, the reflected
hole moves into the subband with spin down. Consequently, the
efficiency of Andreev reflection is determined by the number of
conducting channels in a subband with a smaller value of the Fermi
momentum. As a result, the subgap conductance of an F/S contact
decreases with the increase of the polarization of a ferromagnet
\cite{Beenakker}.

Effects of spin filtering \cite{Martin}, \cite{Barash}, \cite{Vod}
and spin mixing \cite{Fogel} are manifested in the dependence of
moduli and phase shifts of the amplitudes of electron states on
the Fermi surface reflected from $r_{\,\alpha}$
($r_{\,\alpha}\,=\,
\sqrt{R_{\,\alpha}}\,\,exp\,(i\,\theta_{\,\alpha}^{\,r})\,;\,\,R_{\,\alpha}=1-D_{\,\alpha}$)
and transmitted through a potential barrier  $d_{\,\alpha}$
($d_{\,\alpha}\,=\,\sqrt{D_{\,\alpha}}\,\,exp\,(i\,\theta_{\,\alpha}^{\,d})$)
on  $\alpha $ ($\alpha $\,=\,$ \uparrow, \downarrow $  is the spin
index). These effects are the consequence of the presence of the
exchange field in a ferromagnet as well.

The possibility to study the influence of the spin mixing effect
on the $I\,-\,V$ characteristics of superconducting weak links
containing a magnetically active interface appeared after the
boundary conditions (BCs) for the quasiclassical GF were obtained.

In paper \cite{Sauls1}, BCs for the quasiclassical GF for two
metals in contact via a magnetically active interface in terms  of
an interface scattering matrix were derived. These equations were
solved for a junction in the tunneling limit \cite{Sauls1} and for
a contact of a superconductor with a ferromagnetic insulator
\cite{Sauls2}. In paper \cite{Fogel}, BCs for the retarded and
advanced quasiclassical GFs were obtained  in terms of Riccati
amplitudes \cite{Eschrig}, \cite{Shelankov}. In paper
\cite{Sauls3}, BCs  in terms  of Riccati amplitudes were obtained
for the nonequilibrium quasiclassical GF.

The equations, obtained in papers  \cite{Fogel} and \cite{Sauls3},
were solved for magnetically active interfaces with finite
transmission (for SFS \cite{Barash}, \cite{Fogel}, for NFS
\cite{Sauls3} (N stands for normal metal), for S-FIF-S
\cite{Barash2}). These solutions show that Andreev bound states
appear within the superconducting gap
\cite{Barash},\cite{Fogel},\,\cite{Sauls3},\, and the $0-\pi$
transition in the SFS junction is possible \cite{Barash},
\cite{Fogel}.

In papers \cite{Vod} and \,\cite{Vod2}, quasiclassical equations
of superconductivity for metals with a spin-split conduction band
were derived and BCs for  the temperature quasiclassical GF for
the F/S interface were obtained.  The model interface was the same
as in  \cite{Sauls1}, \cite{Zaitsev}.

The aim of this work is to study the influence of spin-dependent
phases of the amplitudes of the electron states reflected from and
transmitted through a potential barrier on Andreev reflection in a
point  FIS contact.

Calculations  are carried out by the method of quasiclassical GFs
with BCs for GFs obtained in papers \cite{Vod},\,\cite{Vod2}.
Below  the dependence of the Andreev reflection probability on
spin-dependent phase shifts $\theta_{\,\alpha}^{\,d}$ and
$\theta_{\,\alpha}^{\,r}$ will be found and the results of the
numerical calculation of the dependence  $G_{FIS}(V)$ for a
rectangular potential barrier and ferromagnets with high
polarization will be discussed.

 1.\, {\bf {Differential conductance of a point FIS contact}}.
In various hybrid F/S structures Andreev reflection is modified.
The reflected hole has some parameters (for example, the velocity
modulus and phase shift) different from those of the incident
electron because it moves in a subband with the opposite spin.
Such spin-discriminating processes due to the exchange field in a
ferromagnet lead to the formation of Andreev bound states inside
the gap \,\cite{Barash},\,\cite{Fogel}.

The enegy of Andreev bound states depends on the spin index
\cite{Barash},\,\cite{Fogel}. As a result, the spectral density of
condauctance $ G_{FIS} $ of the FIS contact at zero voltage is no
longer  a symmetrical function of energy $\varepsilon$. The
condition of the time reversal invariance has the form
$G_{FIS}(\varepsilon,\,\alpha) $\,=\,$
G_{FIS}(-\,\varepsilon,\,-\,\alpha) $.  The generalization of the
conductance $ G_{FIS}(V)$ \,\, \cite{Vod}, \cite{Tinkham} for this
case results in the following expression for  $ G_{FIS}(V)$:
$$
G_{FIS}(V)=\frac{e^{\,2} A}{32\pi^{\,2}\,T} \sum_\alpha\,\rm {Tr}
\left[ \int\frac{d{\bf p}_\|}{\,(2\,\pi)^{\,2}}
\int\limits_{-\infty}^{\infty} d\varepsilon\,\times \right.
$$
$$
\frac{1}{\coth^2(\frac{\varepsilon-eV\,\hat{\tau}_z}{2\,T})}\,\,
[1-\hat{g}_{\,s}^{\,A}\, \tau_z \,\hat{g}_{\,s}^{\,R}
\,\hat{\tau}_{\,z} -\hat{g}_{\,a}^{\,A}\,\hat{\tau}_z
\,\hat{g}_{\,a}^{\,R}\,\hat{\tau}_{\,z}
$$
\begin{equation}
\left. +\hat{\Upsilon}_{\,s}^{\,A}\hat{\tau}_z
\hat{\Upsilon}_{\,s}^{\,R}\,\hat{\tau}_{\,z}-\hat{\Upsilon}_{\,a}^{\,A}
\hat{\tau}_z \hat{\Upsilon}_{\,a}^{\,R}
\hat{\tau}_{\,z}]\right].\label{eq:1}
\end{equation}

In Eq. (\ref{eq:1}), \,$A$\, is the contact area; $\hat{\tau}_z$
is the Pauli matrix; \, $p_\|$ is the momentum in the contact
plane;\, ($\hat{g}_{\,s}$, $\hat{\Upsilon}_{\,s}$) and
\,($\hat{g}_{\,a} $, $\hat{\Upsilon}_{\,a}$) are quasiclassical
retarded (R) and advanced (A) GFs symmetric and antisymmetric\,
with respect to the projection of the momentum \,$ {\bf\hat{p}}$
\, on the Fermi surface on the axis $x$, respectively \cite{Vod}.
Calculations in Eq. (\ref{eq:1}) are to be carried out on the
boundary of any contacting metal.

2. {\bf Finding  GFs and conductance}. Let us assume that the
barrier with the width $d$ is located in the region $ a<x<b
\,\,\,\,(d\,=\,b\,-\,a)$, the superconductor occupies the region
$x>b$, and the ferromagnet occupies the region $x<a$. To find GFs,
for each metal one has to solve quasiclassical equations of
superconductivity for metals with a spin-split conductivity band
simultaneously with their BCs derived in paper \cite{Vod}:
$$
{\rm{sign}}(\hat{p}_{\,x})\frac{\partial}{\partial{\,x}}\,\hat{g}+\frac{1}{2}\,
{\bf
v_\|}\frac{\partial}{\partial{\bf\rho}}(\hat{v}_{\,x}^{-1}\hat{g}+\hat{g}\,\hat{v}_{\,x}^{-1})
+[\hat{K},\,\hat{g}]_- = 0,
$$
$$
\hat{K}=\,-\,i\hat{v}_{\,x}^{-\frac{1}{2}}(i\varepsilon_n
\hat{\tau}_z+
\hat{\Delta}-\hat{\Sigma})\hat{v}_{\,x}^{-\frac{1}{2}} -
i(\hat{p}_{\,x}-\hat{\tau}_x\hat{p}_{\,x}\hat{\tau}_x)/2,
$$
\begin{equation}\label{eq:2} [a,\,b]_{-} = ab - ba . \qquad
\end{equation}

In this section, $\varepsilon_n = (2n + 1)\pi T $  is the
Matsubara frequency; $\hat{\Sigma}$ is the self-energy part;
$\hat{g}$  are matrix temperature GFs:
$$ \hat{g}=
  \begin{pmatrix}
    g_{\,\alpha\,\alpha} & f_{\,\alpha\,-\alpha} \\
    f_{-\,\alpha\,\alpha}^{\,+}&\,-\, g_{-\,\alpha\,-\alpha}
  \end{pmatrix}, \,\,
 \hat{g}=\left\{\begin{split}\hat{g}_>\qquad  \hat{p}_{\,x}>0,\\
\hat{g}_< \qquad  \hat{p}_{\,x}<0
\end{split}\right. .$$
Moreover,
$$
 \hat{\Delta} = \begin{pmatrix}
   0& \Delta \\
    -\Delta^* &0
  \end{pmatrix},\qquad
  \;\hat{p}_{x}=
  \begin{pmatrix}
   p_{\,x,\,\,\alpha }& 0 \\
   0 & p_{x,\,{-\alpha}}
  \end{pmatrix},$$
where $\Delta$  is the order parameter,\, and $ p_{x}$  is the
projection of the momentum  on the Fermi surface on the axis $x$.
Matrices  \,$\hat{v}$ have the same structure as \,\,$
\hat{p}_{\,x}$ .

BCs for the specular reflection of electrons from the boundary:
 $ p_\parallel $ = $p_{\downarrow}\sin\vartheta_\downarrow
$ =
 $p_{\uparrow}\sin\vartheta_\uparrow =
p_S\sin\vartheta_S $,   have the form  \cite{Vod}:
$$
(\hat{\tilde{g}}_a^S)_d = (\hat{\tilde{g}}_a^F)_d,\;\;
(\hat{\tilde{\Upsilon}}_a^S)_d = (\hat{\tilde{\Upsilon}}_a^F)_d,
$$
$$
(\sqrt{\hat{R}_\alpha}-\sqrt{\hat{R}_{-\alpha}})(\hat{\tilde{\Upsilon}}_a^+)_n
= \alpha_3(\hat{\tilde{g}}_a^-)_n,
$$
$$
(\sqrt{\hat{R}_\alpha}-\sqrt{\hat{R}_{-\alpha}})(\hat{\tilde{\Upsilon}}_a^-)_n
= \alpha_4(\hat{\tilde{g}}_a^+)_n,
$$
$$
-\hat{\tilde{\Upsilon}}_s^- =
\sqrt{\hat{R}_\alpha}(\hat{\tilde{g}}_s^+)_d+\alpha_1(\hat{\tilde{g}}_s^+)_n,
$$
\begin{equation}\label{eq:3}
-\hat{\tilde{\Upsilon}}_s^+ =
(\hat{R}_\alpha)^{-\frac{1}{2}}(\hat{\tilde{g}}_s^-)_d+\alpha_2(\hat{\tilde{g}}_s^-)_n,
\end{equation}
where\; $\hat{\tilde{g}}_{a(s)}^{\pm}
 = 1/2\,[\,{\hat{\tilde{g}}}_{a(s)}^S \pm {\hat{\tilde{g}}}_{a(s)}^F\,]$. \;
Functions $\hat{\tilde{\Upsilon}}_{a(s)}^{\pm}$ are determined
analogously. The index $ d $ denotes the diagonal and  $ n $ the
nondiagonal part of the matrix $ \hat{T}_{d(n)} = 1/2\,[ \,
\hat{T} \pm \tau_z\hat{T}\tau_z \,]$. Coefficients $ \alpha_i$
are:
$$
\alpha_{1(2)} = \frac{1+\sqrt{{R_{\,\uparrow}}R_{\,\downarrow}}
\mp \sqrt{{D_{\,\uparrow}}D_{\,\downarrow}}}{\sqrt{R_{\,\uparrow}}
+ \sqrt{R_{\,\downarrow}}},\;\;
$$
$$
\alpha_{3(4)} = 1-\sqrt{{R_{\,\uparrow}}R_{\,\downarrow}}\, \pm
\sqrt{{D_{\,\uparrow}}D_{\,\downarrow}}\;).\;
$$

One can exclude GFs $\hat {\tilde{\Upsilon}}_{\,a}^{\,F}$ and
$\hat {\tilde{\Upsilon}}_{\,a}^{\,S} $ from these relations and
obtain a system of BCs only for the GF $\hat{\tilde{g}}$\,
\cite{Vod2}:
\begin{equation}\label{eq:4}
\begin{split}
{\hat{\tilde{g}}}_a^+ {\widehat{b}}_1 + {\widehat{b}}_2
{\hat{\tilde{g}}}_a^+ + {\hat{\tilde{g}}}_a^- {\widehat{b}}_3 +
{\widehat{b}}_4 {\hat{\tilde{g}}}_a^-
={\widehat{b}}_3-{\widehat{b}}_4 ,\\
{\hat{\tilde{g}}}_a^- {\widehat{b}}_1 + {\widehat{b}}_2
{\hat{\tilde{g}}}_a^- + {\hat{\tilde{g}}}_a^+ {\widehat{b}}_3 +
{\widehat{b}}_4 {\hat{\tilde{g}}}_a^+
={\widehat{b}}_1-{\widehat{b}}_2.
\end{split}
\end{equation}
Matrices $ {\widehat{b}}_i $ in Eq. (\ref{eq:4}) are:
\begin{equation}\label{eq:5}
\begin{array}{cc}
{\widehat{b}}_1 = \hat {\tilde{\Upsilon}}_s^+
{\hat{\tilde{g}}}_s^- + \hat {\tilde{\Upsilon}}_s^-
{\hat{\tilde{g}}}_s^+ ,  & {\widehat{b}}_2 = {\hat{\tilde{g}}}_s^+
\hat {\tilde{\Upsilon}}_s^- +
{\hat{\tilde{g}}}_s^- \hat{\tilde{\Upsilon}}_s^+ ,\\
{\widehat{b}}_3 = \hat {\tilde{\Upsilon}}_s^+ \hat{\tilde{g}}_s^+
+ {\hat {\tilde{\Upsilon}}}_s^- \hat{\tilde{g}}_s^- , &
{\widehat{b}}_4 = {\hat{\tilde{g}}}_s^+ \hat
{\tilde{\Upsilon}}_s^+ + {\hat{\tilde{g}}}_s^-
\hat{\tilde{\Upsilon}}_s^- .
\end{array}
\end{equation}
GFs \,\,$\hat{\tilde{g}}$ \, are connected with GFs \, being
solutions of Eq. (\ref{eq:2}) by the following relationships
\cite{Vod}:
$$
(\hat{\tilde{g}}_s^S)_n\,=\,(\hat{g}_s^S)_n\,\,
\cos(\,\theta_{\,\alpha}) +i\hat{\tau}_z \,(\hat{g}_a^S)_n
\,\,\sin(\,\theta_{\,\alpha})
$$
$$
(\hat{\tilde{g}}_a^S)_n\,=\,(\hat{g}_a^S)_n\,\,\cos(\,\theta_{\,\alpha})
+i\hat{\tau}_z \,(\hat{g}_s^S)_n\,\, \sin(\,\theta_{\,\alpha})
$$
$$
(\hat{\tilde{g}}_s^F)_n\,=\,(\hat{g}_s^F)_n\,\,\cos(\beta_{\,\alpha}^{\,r})
+i\hat{\tau}_z (\hat{g}_a^F)_n\,\, \sin(\beta_{\,\alpha}^{\,r})
$$
$$ (\hat{\tilde{g}}_a^F)_n\,=\,(\hat{g}_a^F)_n\,\,
\cos(\beta_{\,\alpha}^{\,r}) +i\hat{\tau}_z (\hat{g}_s^F)_n\,\,
\sin(\beta_{\,\alpha}^{\,r})
$$
\begin{equation}\label{eq:6}
\theta_{\,\alpha}=\frac{\theta_{\alpha}^{\,r}-
\theta_{-\alpha}^{\,r}}{2}-
(\theta_{\,\alpha}^{\,d}-\theta_{-\,\alpha}^{\,d});\quad \,\,
\beta_{\,\alpha}^{\,r}=\frac{\theta_{\alpha}^{\,r}-
\theta_{-\alpha}^{\,r}}{2} .
\end{equation}
The explicit form of functions $\hat {\tilde{\Upsilon}}$ is not
needed. These functions are found from BCs. The diagonal parts of
matrices $\hat{\tilde{g}}$ are equal to the corresponding matrices
$\hat{g}$. Equations (\ref{eq:2}) for the ballistic contact are
solved in paper \cite{Vod}, \cite{Zaitsev}. At the boundaries for
$x=b$ and $x=a$ we have:
\begin{equation}\label{eq:7}
\hat{g}_s^S=\hat{g}_0^S+\hat{g}_0^S\,\hat{g}_a^S; \quad
\hat{g}_s^F=\hat{g}_0^F-\hat{g}_0^S\,\hat{g}_a^F.
\end{equation}
Matrices $ \hat{g}_0 $ are values of GFs $ \hat{g} $ away from the
boundary:
\begin{gather}
\hat{g}_0^F\,=\,{\rm
{sign}}(\varepsilon_n)\,\hat{\tau}_z \label{eq:8}\\
\hat{g}_0^S  = g_0^S \hat{\tau}_z + (\hat{g}_0^S)_n =
\frac{1}{\sqrt{\varepsilon_n^2 + |\Delta|^2}}
\begin{pmatrix}
  \varepsilon_n & -i\Delta \\
  i\Delta^* & -\varepsilon_n
\end{pmatrix}.\nonumber
\end{gather}
After the substitution of functions $\hat{\tilde{g}}_s^F$ and
$\hat{\tilde{g}}_s^S$, expressed via $\hat{\tilde{g}}_a^F$ and
$\hat{\tilde{g}}_a^S$ by Eq. (\ref{eq:7}), in the system of BCs
Eq.(\ref{eq:4}) and their solution in the linear approximation
with respect to the functions $\hat{\tilde{g}}_a^S$ è
$\hat{\tilde{g}}_a^F$, we find the function $\hat{\tilde{g}}_a^F$:
\begin{gather}
\hat{\tilde{g}}_a^F=
-\,\frac{\sqrt{D_{\uparrow}D_{\downarrow}}\,\hat{\tau}_z\,(\hat{g}_0^S)_n\,
  }{Z} \nonumber\\
Z=(1-\sqrt{R_\uparrow R_\downarrow})\,[g_0^S  \,
\cos(\theta_{\,\alpha})+i\, \sin(\theta_{\,\alpha})] \label{eq:9}\\
+\,(1+\sqrt{R_\uparrow
R_\downarrow})\,{\rm{sign}}(\varepsilon_n)\,
[\cos(\theta_{\,\alpha})\,+i\,g_0^S \,
\sin(\theta_{\,\alpha})].\nonumber
\end{gather}
From Eqs. (\ref{eq:3}) and (\ref{eq:4}) we find the rest functions
necessary to calculate conductance Eq. (\ref{eq:1}) and calculate
conductance at the ferromagnet side.

After carrying out the analytical continuation in these functions
(substitution $i\,\varepsilon_n $\, for\, $ \varepsilon \pm
i\delta \,$ for retarded and advanced GFs, respectively), we
obtain the expression for the conductance $\sigma_{\,F/S}(V)$:
\[
\sigma_{\,F/S}(V)=\frac{\,e^{\,2}\,A}{\pi}\int\frac{d{\bf
p}_{\|}\,}{\,(2\,\pi)^{\,2}}\left\{\,\,
\int\limits_{|\Delta|}^\infty\frac{d\,\varepsilon}{2\,T} \left[
\frac{1}{\cosh^{\,2}(\frac{\varepsilon+eV}{2\,T})}\,\right.\notag
\right.
\]
\[
\left.  +\frac{1}{\cosh^{\,2}(\frac{\varepsilon-eV}{2\,T})}
\right]\frac{\varepsilon\,\xi^{\,R}(D_{\uparrow}+D_{\downarrow})+
\varepsilon\,(\varepsilon-\xi^{\,R})D_{\uparrow}\,D_{\downarrow}}
{Z_\Uparrow}\,\,+ \notag
\]
$$
\int\limits_0^{|\Delta|}\frac{d\,\varepsilon}{2\,T} \left[
\frac{1}{\cosh^{\,2}(\frac{\varepsilon+eV}{2\,T})}+
\frac{1}{\cosh^{\,2}(\frac{\varepsilon-eV}{2\,T})} \right]
\left.\frac{D_{\uparrow}\,D_{\downarrow}\,
|\Delta|^{\,2}}{Z_\Downarrow}\right\}
$$
$$
Z_\Uparrow=[\varepsilon\,(1-W)+ \xi
(1+W)]^{\,2}+4\,W\,|\Delta|^2\sin^2(\theta_{\,\alpha})\nonumber
$$
$$
Z_\Downarrow =[1+2W\cos(2\,\theta_{\,\alpha})+W^{\,2}]|\Delta|^2
-4W\,\varepsilon^{2}\cos(2\,\theta_{\,\alpha})\nonumber
$$
\[
-\,\frac{16 W\,(|\Delta|^2-\varepsilon^2)
\,\varepsilon^{2}\,\sin^2(2\,\theta_{\,\alpha})}
{[1+2W\,\cos(2\,\theta_{\,\alpha})+W^{\,2}]\,|\Delta|^2
-4W\,\varepsilon^{2}\,\cos(2\,\theta_{\,\alpha})}\nonumber
\]
\begin{equation}\label{eq:10}
 W\,=\,\sqrt{R_{\uparrow}R_{\downarrow}};\qquad\xi\,=\,\sqrt{\varepsilon^2
 \,-\,|\Delta|^2}.
\end{equation}
At $\theta_{\,\alpha}\,=\,0$ the expression for conductance
obtained in paper \cite{Vod} follows from Eq. (\ref{eq:10}). In
the case of nonmagnetic metal, when $ D_\uparrow = D_\downarrow $
this expression is the same as that obtained in paper
\cite{Zaitsev}, and for $D\,=\,1/(1+Z^2)$ this expression is the
same as that obtained in paper \cite{Tinkham}.

3.\, {\bf {Andreev reflection}}. The quasiclassical GFs entering
Eq. (\ref{eq:8}) enable the conclusion that
\begin{multline}\label{eq:11}
[1-\hat{g}_{\,s}^{\,A}\, \tau_z \,\hat{g}_{\,s}^{\,R}
\,\hat{\tau}_{\,z} -\hat{g}_{\,a}^{\,A}\,\hat{\tau}_z
\,\hat{g}_{\,a}^{\,R}\,\hat{\tau}_{\,z}
 +\hat{\Upsilon}_{\,s}^{\,A}\hat{\tau}_z
\hat{\Upsilon}_{\,s}^{\,R}\,\hat{\tau}_{\,z}\\
-\hat{\Upsilon}_{\,a}^{\,A} \hat{\tau}_z
\hat{\Upsilon}_{\,a}^{\,R}
\hat{\tau}_{\,z}]=4[-\hat{\tilde{g}}_{\,a}^{\,A}\,\hat{\tau}_z
\,\hat{\tilde{g}}_{\,a}^{\,R}\,\hat{\tau}_{\,z}]\sim \hat{1}.
\end{multline}

Now, the comparison of the form of under-gap conductances in Eq.
(\ref{eq:1}) and that of the corresponding Eq. (25) in  paper
\cite{Tinkham} shows that the matrix elements
$(\hat{\tilde{g}}_{\,a}^{\,R})^F$ and
$(\hat{\tilde{g}}_{\,a}^{\,A})^F$ are the amplitudes of the
Andreev reflection probability
$\tilde{a}(\varepsilon,\,\theta_{\,\alpha})$ in FIS contacts for
energies less than $|\Delta|$
$(\varepsilon^{\,2}\,<\,|\Delta|^{\,2})$. Let us assume that
$\tilde{a}(\varepsilon,\,\theta_{\,\alpha})$ are matrix elements
of $(\hat{\tilde{g}}_{\,a}^{\,R})^F$.
\begin{equation}\label{eq:12}
 \tilde{a}(\varepsilon,\,\theta_{\,\alpha})= \frac{\sqrt{D_{\uparrow}D_{\downarrow}}\,\,
 \Delta}{Z}\,=\,a(\varepsilon,\,\theta_{\,\alpha})\,e^{\,-\,i\,\beta_{\,\alpha}^{\,r}}
\end{equation}
$$
Z=(1-\sqrt{R_\uparrow R_\downarrow})[\varepsilon \,
\cos(\theta_{\,\alpha})-\sqrt{|\Delta|^2-\varepsilon^2}
\sin(\theta_{\,\alpha})]
$$
$$
+i\,(1+\sqrt{R_\uparrow R_\downarrow})
[\sqrt{|\Delta|^2-\varepsilon^2}\cos(\theta_{\,\alpha})+
\varepsilon \, \sin(\theta_{\,\alpha})].
$$
The presence of the imaginary part in functions
$a(\varepsilon,\,\theta_{\,\alpha})$  means that Andreev
reflection is accompanied by the phase shift. The Andreev
reflection probability $A(\varepsilon,\,\theta_{\,\alpha})$\quad
($A(\varepsilon,\,\theta_{\,\alpha})\,=
\,\tilde{a}(\varepsilon,\,\theta_{\,\alpha})\,\tilde{a}^{\,*}(\varepsilon,\,\theta_{\,\alpha})$)
is:
\begin{gather}
    A(\varepsilon,\,\theta_{\,\alpha})\,=\,\frac{D_{\uparrow}\,D_{\downarrow}\,|\Delta|^{\,2}}
    {Z}\label{eq:13}\\
    Z\,=\,[1-\sqrt{R_\uparrow R_\downarrow}]^{\,2}\,|\Delta|^{\,2}\nonumber\\
    +\,4\,\sqrt{R_\uparrow R_\downarrow}\,\,[\sqrt{|\Delta|^{\,2}\,-\,\varepsilon^{\,2}}\,\cos(\,\theta_{\,\alpha})\,
    +\,\varepsilon\,\,\sin(\,\theta_{\,\alpha})]^{\,2}\nonumber.
\end{gather}
It follows from this equation that: (1) in terms of paper
\cite{Fogel} spin-mixing angle $\Theta $ for FIS contact is equal
to $\theta_{\,\alpha}$ (for SFS and NFS contacts $\Theta
$\,=\,$\theta_{\,\uparrow}^{\,r}-\theta_{\,\downarrow}^{\,r}$\,=\,
$\theta_{\,\uparrow}^{\,d}-\theta_{\,\downarrow}^{\,d}$
\cite{Barash},\cite{Fogel},\,\cite{Sauls3}); (2) for
$\theta_{\,\alpha}\,<\,0$ the Andreev reflection probability of
the electron excitation with the spin projection $\alpha$ is
larger than that of the hole excitation; for
$\theta_{\,\alpha}\,>\,0$ the Andreev reflection probability of
the hole excitation with the spin projection $\alpha$ is larger
than that of the electron excitation; (3) the Andreev reflection
probability has maxima at $\varepsilon\,=\,\varepsilon_{\,b}$ (at
the values of the energy of electron (hole) excitations
corresponding to the energy levels of Andreev surface bound
states)
\begin{equation}\label{eq:14}
\epsilon_{\,b}=\left\{\begin{split}
\varepsilon\,=\,|\Delta|\,\cos(\,\theta_{\,\alpha})\quad
\mbox{for}\quad
\theta_{\,\alpha}\,<\,0, \\
\varepsilon\,=-\,|\Delta|\,\cos(\,\theta_{\,\alpha}) \quad
\mbox{for} \quad \theta_{\,\alpha}\,>\,0.
\end{split} \right.
\end{equation}
\begin{figure}[h]
\centering \resizebox{0.5
\textwidth}{!}{\includegraphics{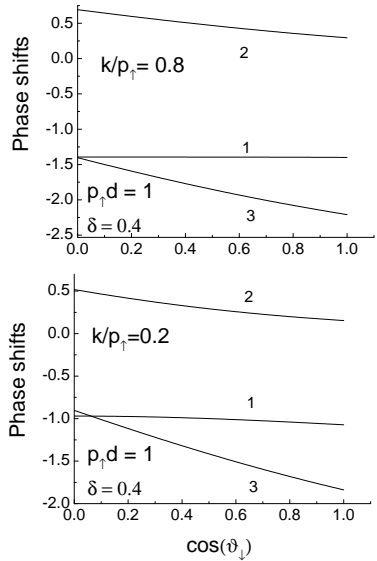}} \caption{Dependence of
the phase shifts of reflection and transmission amplitudes on
$\cos(\,\vartheta_{\,\downarrow})$. Lines with numbers 1,\,2, and
\,3 depict these dependences  on $\cos(\,\vartheta_{\,\downarrow})
$:\,\,$\theta_{\,\uparrow}$;
($\theta_{\,\uparrow}^{\,d}-\theta_{\,\downarrow}^{\,d}$) and
($\theta_{\,\uparrow}^{\,b}-\theta_{\,\downarrow}^{\,b}$),
 respectively.}
\end{figure}
\begin{figure}[h]
\centering
\resizebox{0.5\textwidth}{!}{\includegraphics{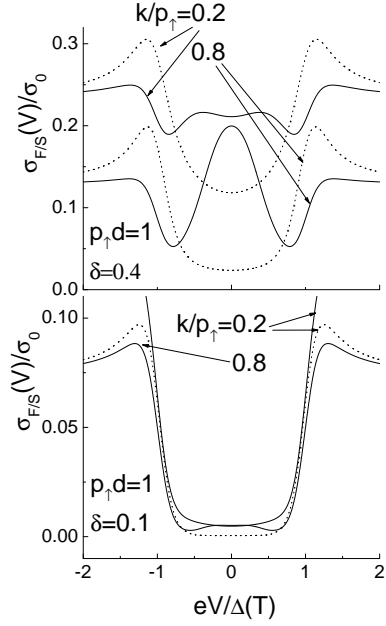}} \caption
{Dependence of the normalized conductance
$\sigma_{\,F/S}(V)/\sigma_0$  from Eq. (\ref{eq:10}) on the
applied voltage for different values of the polarization of a
ferromagnet $ \delta\,$=$\,p_{\,\downarrow}/p_{\,\uparrow}$ at the
ratio $\Delta_{d}(T)/2T$ \,=$\,6 $.}
\end{figure}
    Below the results of the numerical calculations of phase shifts
and conductance are presented.  In the numerical calculations the
relation between Fermi momenta of contacting metals was the
following: $p_{\,S}\,=\,(p_{\,\uparrow}\,+\,p_{\,\downarrow})/2$.
Calculations are carried out for a rectangular barrier with the
height $U$ counted off the bottom of the conduction band of a
superconductor; [$\chi(x)$ is the wave function of an electron in
an isolator,
\,$\chi(x)\,=\,C_1\,\exp(\,\gamma\,x)\,+\,C_2\,\exp(-\,\gamma\,x);\,\,
\gamma\,=\,\sqrt{k^2\,+\,p_{\,\parallel}^{\,2}};\,k^2\,=\,2m_b(U\,-\,E_F^S)\,$;
$E_F^S$   is the Fermi energy of a superconductor, $m_b$ is the
mass of an electron in a barrier].   In this case the expressions
for $ \theta_{\,\alpha}^{\,d}$ and $ \theta_{\,\alpha}^{\,r}$ have
the following form:
$$
    \theta_{\alpha}^{\,d}=\widetilde{\theta}_{\,\alpha}^{\,d}+i(p_{\,x,\,\alpha}^{\,F}\,a-p_{\,x}^{\,S}\,b);
    \quad \theta_{\,\alpha}^{\,r}\,=\,\widetilde{\theta}_{\,\alpha}^{\,r}\,+\,2ip_{\,x,\,\alpha}^{\,F}\,a
$$
    \begin{equation}\label{eq:15}
    \widetilde{\theta}_{\,\alpha}^{\,d}\,=\,\arctan\left(\frac{(p_{\,x,\,\alpha}^{\,F}\,p_{\,\,x}^{\,S}-
    \gamma^{\,2})\tanh(\gamma \, d)}
    {\gamma\,(p_{\,x,\,\alpha}^{\,F}\,+\,p_{\,x}^{\,S})} \right)
\end{equation}
    $$
    \widetilde{\theta}_{\,\alpha}^{\,r}\,=\,\arctan \left(\frac{2\,
    \gamma \,
    p_{\,x,\,\alpha}^{\,F}\,[\gamma^{\,2}+(p_{\,x}^{\,S})^{\,2}]
    \tanh(\gamma \,d)}{Z} \right)
    $$
    $$
    Z=\gamma^{\,2}\,[(p_{\,x}^{\,S})^{\,2}-(p_{x,\,\alpha}^{\,F})^{\,2}] +
    [\gamma^{\,2}-p_{\,x}^{\,S}\,p_{x,\,\alpha}^{\,F}]^{\,2}\tanh^{2}(\gamma
    d),
    $$
so that the angle \,\,$\theta_{\,\alpha}$\,\,\,\,
$[\theta_{\,\alpha}=(\theta_{\alpha}^{\,r}-
\theta_{-\alpha}^{\,r})/2-
(\theta_{\,\alpha}^{\,d}-\theta_{-\,\alpha}^{\,d})]\,=\,(\widetilde{\theta}_{\alpha}^{\,r}-
\widetilde{\theta}_{-\alpha}^{\,r})/2-
(\widetilde{\theta}_{\,\alpha}^{\,d}-\widetilde{\theta}_{-\,\alpha}^{\,d})$
does not depend on the location of the barrier.

Figure 1  shows the  dependences of the phase shifts on
$\cos(\,\vartheta_{\,\downarrow})$. All angles are connected by
specular reflection $ p_\parallel $ =
$p_{\downarrow}\sin\vartheta_\downarrow $ =
$p_{\uparrow}\sin\vartheta_\uparrow = p_S\sin\vartheta_S $.

The phase shift $ \theta_{\,\uparrow }$ slowly decreases as the
polarization of the ferromagnet $\delta $ decreases [from (-1.3)
at $\delta $\,=\,0.05 to (-1.5) at $\delta $=0.5 ($p_\uparrow
d\,=\,1;\, k/p_\uparrow\,=\,0.8 $) ] and [from (-0.7) at
$\delta$=0.05 to (-1.2) at $\delta$\,=\,0.5 ($p_\uparrow
d\,=\,1;\, k/p_\uparrow\,=\,0.2 $ )]. It means that the points
$\epsilon_{\,b}$ approach zero as the polarization decreases and
$k/p_\uparrow$ increases, however, at $k/p_\uparrow\,>\,1$,
$\theta_{\,\uparrow}$ rapidly decreases down to zero. 6.  With the
increasing parameter  $p_{\,\uparrow }\,d$ and other parameters
fixed  (but for $k/p_\uparrow\,<\,1)$ the angle
$\theta_{\,\alpha}$ also tends to $\pi/2$. Note that the
spin-mixing angle  $\theta_{\,\alpha}$ for ferromagnets with large
polarization is practically the same for all electron
trajectories.

The upper panel in Fig. 2 shows the results of the numerical
calculations carried out according to Eq. (\ref{eq:10}) not taking
into account (dashed lines) and taking into account (solid lines)
the phase shift $\theta{\,\alpha}$. The peaks in the dependence of
the conductance on $V $ (Fig. 2, upper panel) correspond to the
motion of the energy levels of Andreev surface bound states
towards each other as the parameter $k/p_\uparrow$ increases.

The lower panel in Fig. 2 shows the suppression of Andreev
reflection due to the reduction of the number of conducting
channels in a subband with a lower value of the Fermi momentum and
the effect of spin filtering.
\begin{figure}[h]
\centering
\resizebox{0.4\textwidth}{!}{\includegraphics{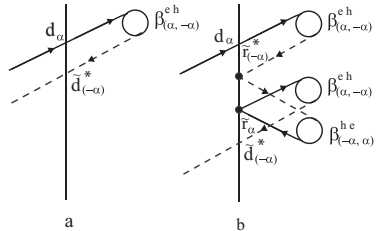}} \caption
{Structure of the diagrams corresponding to Andreev reflection in
the superconductor:  diagram a) one-act process; diagram b)
two-act process. The vertex  $\bigcirc$ is Andreev reflection of
electronlike (solid lines) and holelike (broken lines)
quasipaticles by the pair patential. The vertex $\bullet$ is the
normal reflection of electronlike and holelike quasipaticles by
the barrier potential. When the solid line transforms into the
broken line,   $\bigcirc$ denotes the vertex
$\beta_{\alpha,\,-\alpha}^{\,e\,h} $. When the broken line
transforms into the solid line, $\bigcirc$ denotes the vertex $
\beta_{-\alpha,\alpha}^{\,h\,e}$. Parameters
$d_{\,\alpha},\,\tilde{d}_{\,\alpha},\,r_{\,\alpha}$ and
$\tilde{r}_{\,\alpha}$ are related as follows:
$\tilde{d}_{\,\alpha}\,=\,d_{\,\alpha}\,p_{\,x}^{\,S}/p_{\,x\,\alpha}^{\,F}$;
$\tilde{r}_{\,\alpha}\,=\,-\,r_{\,\alpha}^*\,d_{\,\alpha}/d_{\,\alpha}^*;\,\,
D_{\,\alpha}\,=\,d_{\,\alpha}\,\tilde{d}_{\,\alpha}^{\,*}$\,\,
\cite{Zaitsev}.}
\end{figure}
Andreev surface bound states are formed in a superconductor due to
the interference of electronlike and holelike particles with
different spin-dependent phase shifts. To demonstrate this, let us
consider diagrams in Fig. 3, corresponding to Andreev reflection
of an electron with the spin projection $\alpha$ and the energy
less than $|\Delta|$ ïtransmitted from a ferromagnet into a
superconductor. The amplitude $a(\varepsilon,\,\theta_{\,\alpha})$
is:
\begin{gather}
   a(\varepsilon,\,\theta_{\,\alpha})\,=\,d_{\,\alpha}\,\tilde{d}_{\,-\alpha}^*\,\beta_{\alpha,-\alpha}^{\,e\,h}[1\,+\,
    \tilde{r}_{\,-\alpha}^*\,\tilde{r}_{\,\alpha}\,
    \beta_{\alpha,\,-\alpha}^{\,e\,h}\,\beta_{-\alpha,\,\alpha}^{\,h\,e}\,\nonumber\\
+(\tilde{r}_{\,-\alpha}^*\,\tilde{r}_{\,\alpha}\,
\beta_{\alpha,\,-\alpha}^{\,e\,h}\beta_{-\alpha,\,\alpha}^{\,h\,e})^2+...]=
\,\frac{d_{\,\alpha}\,\tilde{d}_{\,-\alpha}^*\,\beta_{\alpha,\,-\alpha}^{\,e\,h}}
{1-\tilde{r}_{-\alpha}^*\tilde{r}_{\alpha}\,\beta_{\alpha,-\alpha}^{\,e\,h}\beta_{-\alpha,\alpha}^{\,h\,e}}\nonumber\\
=\frac{\sqrt{D_{\alpha}D_{-\alpha}\,p_{\,x\,\alpha}^{\,F}/p_{\,x\,-\,\alpha}^{\,F}}\,\,\,
e^{\,i\,\beta_{\,\alpha}^{\,r}}
\,\,\,\beta_{\alpha,\,-\alpha}^{\,e\,h}}
{e^{\,i\,\theta_{\,\alpha}}\,-\,
e^{\,-i\,\theta_{\,\alpha}}\sqrt{R_{\alpha}R_{-\alpha}}\,
\,\beta_{\alpha,-\alpha}^{\,e\,h}\,\beta_{-\alpha,\alpha}^{\,h\,e}}.\label{eq:16}
\end{gather}
The corresponding probability of Andreev reflection is:
\begin{equation}\label{eq:17}
    A(\varepsilon,\,\theta_{\,\alpha})\,=\,\frac{D_{\alpha}D_{-\alpha}\,p_{\,x,\,\alpha}^{\,F}/p_{\,x,\,-\,\alpha}^{\,F}
\,\,\beta_{\alpha,\,-\alpha}^{\,e\,h}\,\,\beta_{\alpha,\,-\alpha}^{*\,e\,h}}
{1+R_{\alpha}R_{-\alpha}\,\,|\beta_{\alpha,\,-\alpha}^{\,e\,h}|^{\,2}\,|\beta_{-\alpha,\alpha}^{\,h\,e}|^{\,2}-
Q}
\end{equation}
$$
Q=\sqrt{R_{\alpha}R_{-\alpha}}\left[\,\cos(2\,\theta_{\alpha})
[\beta_{\alpha,\,-\alpha}^{\,e\,h}\,\beta_{-\alpha,\,\alpha}^{\,h\,e}+
\beta_{\alpha,\,-\alpha}^{*\,e\,h}\,\beta_{-\alpha,\,\alpha}^{*\,h\,e}]\right.
$$
$$
\left.+\,i\,\sin(2\,\theta_{\alpha}) [
\beta_{\alpha,\,-\alpha}^{*\,e\,h}\,\beta_{-\alpha,\,\alpha}^{*\,h\,e}-
\beta_{\alpha,\,-\alpha}^{\,e\,h}\,\beta_{-\alpha,\,\alpha}^{\,h\,e}]\right]
$$
By comparing formulas (\ref{eq:16},\,\ref{eq:17})  with formulas
(\ref{eq:12},\,\ref{eq:13}) we find the vertices
$\beta_{\alpha,\,-\alpha}^{\,e\,h}$ and
$\beta_{-\alpha,\alpha}^{\,h\,e}$:
\begin{equation}\label{eq:18}
   \beta_{\alpha,\,-\alpha}^{\,e\,h}\,=\,\sqrt{\frac{p_{\,x,\,-\,\alpha}^{\,F}}{p_{\,x,\,\alpha}^{\,F}}}\quad
   \frac{\varepsilon\,-\,i\,\sqrt{|\Delta|^{\,2}-\varepsilon^{\,2}}}{|\Delta|}\,\frac{\Delta}{|\Delta|}
\end{equation}
   $$
   \beta_{-\alpha,\alpha}^{\,h\,e}\,=\,\sqrt{\frac{p_{\,x,\,\,\alpha}^{\,F}}{p_{\,x,\,-\,\alpha}^{\,F}}}\quad
   \frac{\varepsilon\,-\,i\,\sqrt{|\Delta|^{\,2}-\varepsilon^{\,2}}}{|\Delta|}\,\frac{\,\,\Delta^*}{|\Delta|}.
   $$
    It follows from formula  (\ref{eq:17}) that in the absence of the interferential term
$Q$ the probability of Andreev reflection is a constant
(independent of the energy $\varepsilon$) quantity. The
interference of electronlike and holelike particles reflected by
the pair potential and interface results in the formation of
Andreev surface bound states. At  $\theta_{\,\alpha}\,=0\,$
 the maximum in the probability of Andreev reflection is at
$\varepsilon\,=\,\pm\,|\Delta|$ \cite{Tinkham}.  At
$\theta_{\,\alpha}\,=\,\pm\,\pi/2\,$ Andreev surface bound states
with the width $\Gamma$ equal to:
\begin{equation}\label{eq:19}
    \Gamma\,=\,\frac{(1-\sqrt{R{\uparrow}\,R{\downarrow}}\,)\,|\Delta|}
{2\,\sqrt[4]{R{\uparrow}\,R{\downarrow}}}
\end{equation}
are formed at $\varepsilon\,=\,0$  on the Fermi level.  The peak
in the differential conductance of an FIS contact at the zero
voltage may be used to determine the polarization of strong
ferromagnets by comparing experimental data with those calculated
according to formula (\ref{eq:10}).

Thus, in the present paper the ballistic conductance of the point
FIS contact is calculated. The dependence of Andreev surface bound
states on the spin-dependent phase shifts of the electron states
reflected from and transmitted through the potential barrier is
found for the interface with finite transmission. By the example
of a rectangular potential barrier it is shown that these states
are manifested in the peaks of the dependence of the conductance
of the FIS contact on the applied voltage.

I am grateful to G.B. Teitel'baum  and  V.V. Ryazanov for
discussing the resalts of this work.

  The work is supported by the  Russian Foundation for
Basic Research, grant ¹ 06-02-17233.

\end{document}